# On new approaches to the study of quantum nanosystems

N.N.Trunov

trunov@vniim.ru

*D.I.Mendeleyev Institute for Metrology*
*Russia, St.Peterburg, 190005 Moskovsky pr. 19*

(Dated: February 25, 2013)

**Abstract**: Quantum nanosystems are extremely diverse and often very complicated. That's why new methods of a simple description of such systems ensuring the retention of essential pert of information at small numbers of parameters are needed. We consider several approaches to such reduced description.

Besides we warn against using of one new incorrect approach.

### 1. Reduced description

An enormous variety of quantum nanoobjects and nanosystems calls for the development of new approaches to their description and parametrization [1,2]. Corresponding methods should be simple and universal enough, ensuring the retention of essential part of information at small number of parameters. Hereafter we discuss several ways of such reduced description.

A universal effective quantum number $T$ was introduced and calculated for centrally symmetric systems [3]

$$T = n + \tfrac{1}{2} + \varphi(l + \tfrac{1}{2}),$$

where *n* and *l* are the radial and orbital quantum numbers. Though *T* is formally not the exact quantum number, it determines with high accuracy both the level ordering and energy values of bound states. The phase $\varphi$ is a certain functional of a given potential and a smooth function of energy. Many improvements of this approach may be developed [2, 6].

In many cases existing theory does not allow us to connect two properties of nanosystems or this connection is too complicated. Then we can use the study of correlations of properties on a set of similar nanoobjects. [2, 4]. The well known normalized coefficient of correlation *Cov* must be near to unity and the possibility of random evidence small enough if the first property really correlates with the second one.

In such a way the objectivity and the quality of a given form of the Periodic system of the elements as a single whole was successfully investigated [5].

In many cases one of modern methods of interpolation may be used for the reduced description. For example, the two-point Pade' approximation usually allows to retain the essential part of information by means of a small number of coefficients with high accuracy [2]. In different situations we can interpolate in parameter space (e.g. numbers of atoms in clusters [1, 2]) as well as between two known results for limiting values of a characteristic physical parameter [2].

Combining two of the abovementioned methods we get new possibilities for the reduced description.

**2. An incorrect approach**

In this section we'd like to forewarn about an incorrect approach [7] to the study of physical effects in nanosystems. It is stated in Ref. 7 that "the interpretation of quantum mechanics by means of physical variables, i.e. without wave functions, allows to eliminate existing contradictions and to predict new physical effects". It is proposed to use a specific form of the Schroedinger equanion and add some new arbitrary suppositions.

The basic object of the quantum mechanics is the two-point density matrix $\rho$. It has a factorized form

$$\rho(x,x') = \Psi(x)\Psi^*(x')$$

with a Hilbert space vector $\Psi$ - also called the wave function - if and only if $\rho$ is diagonal in this space. Really, such a case often takes place in simple problems. But we cannot add arbitrary interpretation of $\Psi$ since it follows from the fundamental properties of $\rho$. The usual density corresponds to the coinciding arguments: $\rho(x,x)$.

The so-called hydrodynamic form of the Schroedinger equation is based on the following representation

$$\Psi = A \exp\left[\frac{iS}{\hbar}\right].$$

Obviously, if the action $S$ is real, the density is $A^2$

The known transformations lead to formally classical equations for $A^2$ and the flow with the effective potential energy

$$V_{eff} = V - \frac{\hbar^2}{2mA}\Delta A.$$

Such "quantum addition" may be useful for semiclassical calculations, but in no way means a reinterpretation of $\Psi$.

In [7, Ch. 10] a stationary one-dimensional motion in the field of a rectangular barrier is also considered. The known reflection coefficient

$$R(E,0) = \left[\frac{k-q}{k+q}\right]^2$$

for the barrier height $U$, the energy value $E$,

$$k = \frac{\sqrt{2m(E-U)}}{\hbar}, \qquad q = \frac{\sqrt{2mE}}{\hbar}.$$

In this model case $R$ does not depend on the Planck constant $\hbar$.

We can treat the above barrier as a limiting case of the potential barrier

$$V(x) = \frac{U}{1+\exp(-x/\alpha)}$$

when the inclined interval $\alpha$ tends to zero:

$$q\alpha \ll 1.$$

The general expression

$$R(E,\alpha) = \left[\frac{sh\,\pi\alpha(k-q)}{sh\,\pi\alpha(k+q)}\right]^2$$

returns in such a case to the above $R(E, 0)$ and its dependence on $\hbar$ vanishes.

Thus, the statement in Ch. 10 that "such description is unsatisfactory and needs additional suppositions" is incorrect. Correspondingly, all the calculations in following chapters, using such suppositions, are also erroneous.

The wave function of the semiinfinite free motion

$$\Psi = \exp(-ipx), \quad x > 0;$$
$$\Psi = 0, \qquad\qquad x \leq 0$$

is not the eigenfunction of momentum and that's why it has no definite "macroscopic momentum"(as it is stated in [7]). Strongly speaking, such situation must be treated by means of the Schroedinger equation with a source.

The time-independent flow of many particles differs from the sole localized particle. In the last case we have a non-stationary wave packet. It is shown long ago that for most of characteristic parameters (of scattering, e.g.) it is possible to replace such localized wave function by a simple plane wave [8, Ch. X ].

Thus the whole theoretical approach [7] is a mixture of classical, semiclassical and quantum ideas with some new arbitrary assumptions. No correct results may be achieved in this way.